\pdfoutput=1
\documentclass[11pt]{article}

\usepackage[final]{acl}

\usepackage{times}
\usepackage{latexsym}
\usepackage{algorithm}
\usepackage{algorithmic}
\usepackage{makecell}
\usepackage{multirow}
\usepackage{bigstrut}
\usepackage{subfigure}
\usepackage{booktabs}

\usepackage[T1]{fontenc}

\usepackage[utf8]{inputenc}

\usepackage{microtype}

\usepackage{inconsolata}

\usepackage{graphicx}

\title{RealVul: Can We Detect Vulnerabilities in Web Applications with LLM?}

\author{
Di Cao$^{1}$,~Yong Liao$^{1}$\thanks{ \;  Corresponding authors.}, Xiuwei Shang$^{1}$\\
        $^{1}$School of Cyber Science and Technology, University of Science and Technology of China \\ 
        \texttt{\{ishgard, shangxw\}@mail.ustc.edu.cn, yliao@ustc.edu.cn} \\
}

\begin{document}
\maketitle

\begin{abstract}
The latest advancements in large language models (LLMs) have sparked interest in their potential for software vulnerability detection. However, there is currently a lack of research specifically focused on vulnerabilities in the PHP language, and challenges in extracting samples and processing persist, hindering the model's ability to effectively capture the characteristics of specific vulnerabilities. In this paper, we present RealVul, the first LLM-based framework designed for PHP vulnerability detection, addressing these issues. By vulnerability candidate detection methods and employing techniques such as normalization, we can isolate potential vulnerability triggers while streamlining the code and eliminating unnecessary semantic information, enabling the model to better understand and learn from the generated vulnerability samples. We also address the issue of insufficient PHP vulnerability samples by improving data synthesis methods.
To evaluate RealVul's performance, we conduct an extensive analysis using five distinct code LLMs on vulnerability data from 180 PHP projects. The results demonstrate a significant improvement in both effectiveness and generalization compared to existing methods, effectively boosting the vulnerability detection capabilities of these models.

\end{abstract}

\section{Introduction}

Software vulnerabilities present substantial risks to the security and integrity of computer systems, networks, and data\cite{Cybersecurity_Almanac}. In 2023, a staggering 28,902 vulnerabilities were publicly reported in the Common Vulnerabilities and Exposures (CVE) database\cite{Cve}. PHP, recognized as the most prevalent and extensively utilized language in web applications, powers nearly 80\% of the top ten million websites\cite{W3Techs}. This includes widely adopted platforms such as Facebook, Wikipedia, Flickr, and WordPress. Moreover, it has been instrumental in the development of over 3.3 million open-source projects on GitHub\cite{Github}. However, PHP is susceptible to common web security vulnerabilities, including SQL injection and cross-site scripting (XSS). Consequently, the imperative to effectively detect PHP software vulnerabilities has never been more pressing.

Traditional vulnerability detection methods, such as Static Application Security Testing (SAST) tools including CodeQL\cite{CodeQL}, RIPS\cite{RIPS}, SonarQube\cite{SonarQube}, Fortify SCA\cite{fortify}, and Checkmarx\cite{checkmarx}, are often constrained by the comprehensiveness and precision of their rule libraries. This limitation frequently results in a high incidence of false positives and negatives. And as a popular framework, CodeQL does not yet support PHP language. To address these issues, researchers have begun to explore the application of deep learning for vulnerability detection\cite{li2018vuldeepecker, zhou2019devign, zou2019muVulDeePecker, wang2020Funded, li2021sysevr, chakraborty2020Reveal, mirsky2023vulchecker}. These approaches extract code structure information in the form of Data Flow Graph (DFG) and Control Flow Graph (CFG), and input vulnerability samples into deep learning models for training. Concurrently, with the advancement of Large Language Modeling (LLM), studies focusing on the application of LLM to code vulnerability detection have started to surface\cite{fu2022linevul,wang2023defecthunter, sun2024llm4vuln}.

After reviewing the current research on applying deep learning or LLMs to vulnerability detection, we discovered:
(1) As shown in Appendix \ref{sec:appendix_Related}, existing LLM-based methods predominantly rely on vulnerability datasets in C/C++ languages for analysis, leaving a research gap in other language;  
(2) The majority of vulnerability datasets are collected through vulnerability fixes on GitHub\cite{chakraborty2020Reveal, zhou2019devign, nikitopoulos2021crossvul}, which presents certain challenges in data collection. Our practical experience has shown that the samples in these datasets may not correlate appropriately with vulnerabilities; 
(3) Vulnerable code requires suitable preprocessing before being inputted into the model to reduce noise and highlight vulnerability features, but this aspect is often overlooked in existing research.

To address these issues, we propose RealVul, a new snippet-level PHP vulnerability analysis framework. First of all, RealVul extracts the real-world vulnerability dataset by identifying potential vulnerability trigger points from real-world projects, analyzing control flow and data flow, and proper data preprocessing methods. Then RealVul generates large semi-synthetic vulnerability dataset from real-world vulnerability dataset and projects. RealVul use this semi-synthetic dataset to fine-tune different LLMs including CodeT5\cite{wang2021codet5}, CodeT5+\cite{wang2023codet5+}, StarCoder2\cite{lozhkov2024starcoder}, and CodeLlama\cite{roziere2023codellama}. 

We evaluate RealVul on CWE-79 (XSS) and CWE-89 (SQL Injection), comparing RealVul with existing approaches. The results demonstrate that RealVul achieves reliable generalization performance while ensuring effectiveness, making our framework suitable for detecting PHP vulnerabilities in real-world projects.

Our main contributions are as follows:
\begin{itemize}
    \item \textbf{LLM-based PHP Vulnerability Detection Framework.} To the best of our knowledge, RealVul is the first LLM-based framework to extract vulnerability dataset and detect PHP software vulnerabilities. It significantly enhances the ability of LLMs to detect vulnerabilities and implements scalable vulnerability detection based on robust generalization performance. (sec.\ref{sec:method})
    \item \textbf{Dataset Collection and Preprocessing.} Different from previous dataset collection methods based on vulnerability repair, we extracted our new RealVul dataset from real-world projects by localizing potential vulnerability and program slicing, and we performed appropriate data preprocessing on it. This allows the model to perform better in the task of vulnerability detection. (sec.\ref{sec:Vulnerability_Candidate_Detection} and sec.\ref{sec:Data_Preprocessing})
    \item \textbf{Data Synthesis.} We present a new data synthesis method and generate a large semi-synthetic vulnerability dataset by inserting pure vulnerability samples into real-world projects devoid of vulnerabilities, which allows programming languages like PHP that lack sufficient vulnerability datasets to abtain enough samples for model training. (sec.\ref{sec:data_Synthesis})
    \item \textbf{Evaluations and Findings.} We conduct tests on real-world vulnerability dataset to comprehensively compare and evaluate the vulnerability detection capabilities of RealVul with existing methods. Our research underscores the weakness of existing dataset and the importance of proper preprocessing. (sec.\ref{sec:Evaluation})
\end{itemize}

\section{Related Works}

In this section, we provide an overview of previous works related to vulnerability detection approches and the corresponding datasets. We primarily summarize the datasets and approaches used in Appendix \ref{sec:appendix_Related}, as shown in Table \ref{tab:vulnerability_datasets} and Table \ref{tab:exsiting_approches}.

\begin{figure}[t]
  \includegraphics[width=\columnwidth]{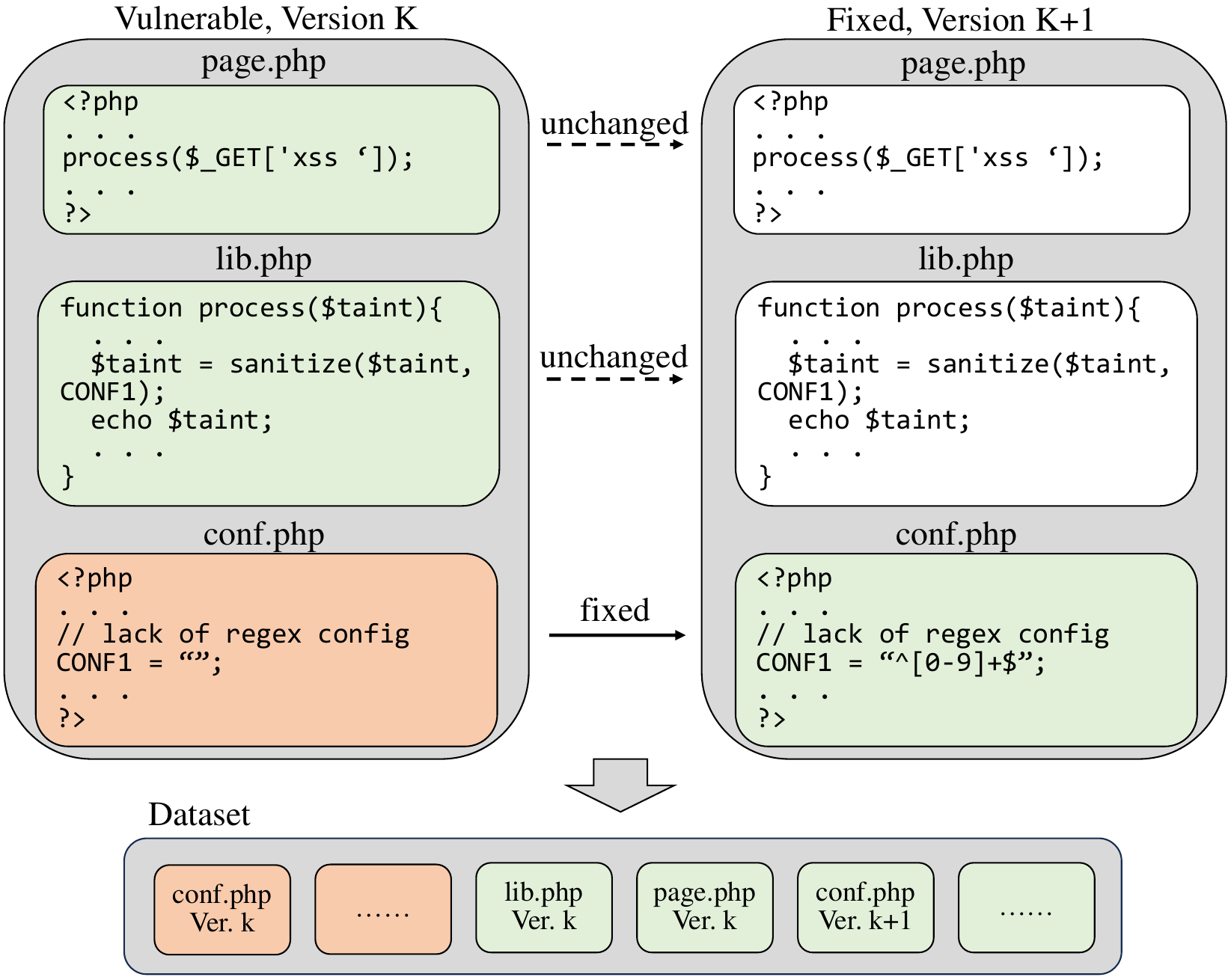}
  \caption{In the case of using vulnerability repair to build a dataset, the green part will be considered secure, and the red part will be considered vulnerable. }
  \label{fig:vuln_fix_example}
\end{figure}

\subsection{Existing Datasets}\label{sec:Existing Datasets}

Previous studies have proposed numerous vulnerability datasets, which can be broadly categorized into two types: synthetic datasets and datasets derived from real project. 

Synthetic datasets are simplified and isolated, and while they contain accurate labels, they lack noise and contextual information, and fail to fully encapsulate the complexity of real-world vulnerabilities. For example, SARD \cite{sard} and Juliet \cite{juliet2013} are overly simplified and do not accurately represent the vulnerabilities that may be encountered in practical applications.

To address the limitations of synthetic datasets, some researchers have suggested collecting data based on vulnerability repair. For vulnerability datasets\cite{zhou2019devign, chakraborty2020Reveal, fan2020Bigvul, D2A, nikitopoulos2021crossvul, chen2023diversevul} derived from real projects, the common approach is to collect vulnerabilities and their corresponding fixes. While this may seem logical, previous research\cite{croft2023data_quality} has indicated that this data collection method still has issues with accuracy, uniqueness, and other aspects.

For instance, consider a potential PHP vulnerability fix, as illustrated in Figure \ref{fig:vuln_fix_example}. 
The execution path of the vulnerability passes through three functions. When extracting samples, this method will split the execution path of the vulnerability into four samples based on whether the function has been fixed. During the model training process, the correlation between samples will be ignored, making it difficult for the model to identify the vulnerability based only on this dataset.

Furthermore, the automatic collection of vulnerability datasets based on vulnerability repair may introduce additional issues, such as including unknown vulnerabilities in the code, which can negatively impact model performance. In our view, vulnerability detection and vulnerability code repair are two different tasks in different stages of vulnerability management. Therefore, we need to take a new approach to extract vulnerability datasets.

\subsection{Existing Approaches}\label{Existing Approches}

Early research primarily employed deep learning models such as RNN and GNN for analysis. Token-based methods transform code into tokens for examination. $\mu$VulDeePecker\cite{zou2019muVulDeePecker} introduces the concept of code attention and utilizes the Building-block BLSTM network. SySeVR\cite{li2021sysevr} incorporates semantic information into the vulnerability syntax candidate SyVCs to generate SeVCs and tests them on models such as BRNN and BGRU.

Graph-based methods transform code into graph structures for examination. Devign\cite{zhou2019devign} and Reveal\cite{chakraborty2020Reveal} leverages the Code Property Graph (CPG) proposed by Yamaguchi et al.\cite{yamaguchi2014CPG} to construct vulnerability prediction model. LineVD\cite{hin2022linevd} utilizes Program Dependency Graph (PDG) to achieve more precise vulnerability localization. VulChecker further \cite{mirsky2023vulchecker} employed ePDG and S2V to further capture the correlation between vulnerability codes.

With the advancements in natural language processing, models such as CodeBERT\cite{feng2020codebert} have demonstrated remarkable code comprehension and generation capabilities, leading more researchers to use them for vulnerability analysis. LineVul\cite{fu2022linevul} employs CodeBERT as its core and implements fine-grained vulnerability classification and localization based on the attention mechanism. DiverseVul\cite{chen2023diversevul} uses models like RoBERTa\cite{liu2019roberta}, GPT-2\cite{radford2019GPT2}, and CodeT5\cite{wang2021codet5} to analyze vulnerability detection capabilities. However, these works are overly reliant on existing datasets and lack reasonable preprocessing of vulnerability code, such as code slicing and duplicate removal. They directly analyze vulnerability code samples, resulting in unreliable analysis of unknown projects\cite{chen2023diversevul}. Recently, the rapid development of large language models has triggered disruptive changes in related fields. It performs outstanding in code-intensive fields such as code synthesis \cite{HowEffective_ISSTA_2023,jiangICSE2023} and automated programming assistance \cite{Leung2023ase,wei2023copiloting}, making its potential in software vulnerability detection obvious.

In RealVul, we process samples reasonably by locating vulnerability triggers, code slicing, irrelevant information permutation, and removing similar samples, and use more advanced code LLM for analysis to achieve superior performance.

\begin{figure*}[t]
  \centering
  \scalebox{0.85}{
  \includegraphics[width=\textwidth]{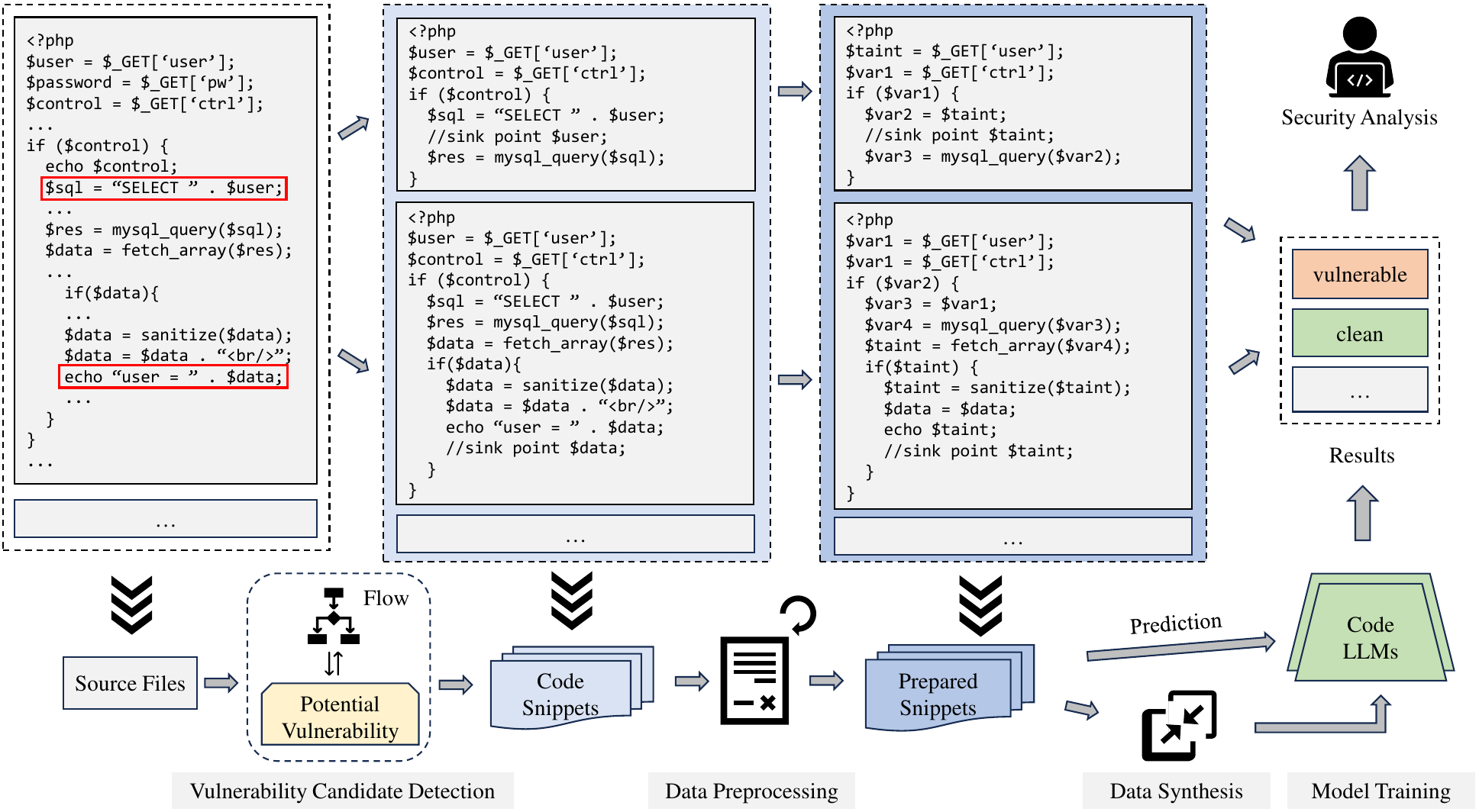}
  }
  \caption{RealVul architecture overview. }
  \vspace{-2ex}
  \label{fig:framework}
\end{figure*}

\section{Method}\label{sec:method}
This section elucidates the design rationale and the architecture of our proposed approach, RealVul. For each type of CWE, RealVul employs distinct strategies for sample selection and processing, and trains separate models for analysis. Our methodology is expounded upon from three perspectives: sample selection, data preprocessing, and model training. Figure \ref{fig:framework} illustrates the architecture of our approach.

\subsection{Vulnerability Candidate Detection}\label{sec:Vulnerability_Candidate_Detection}
\begin{figure*}[t]
  \centering
  \scalebox{0.85}{
  \includegraphics[width=\textwidth]{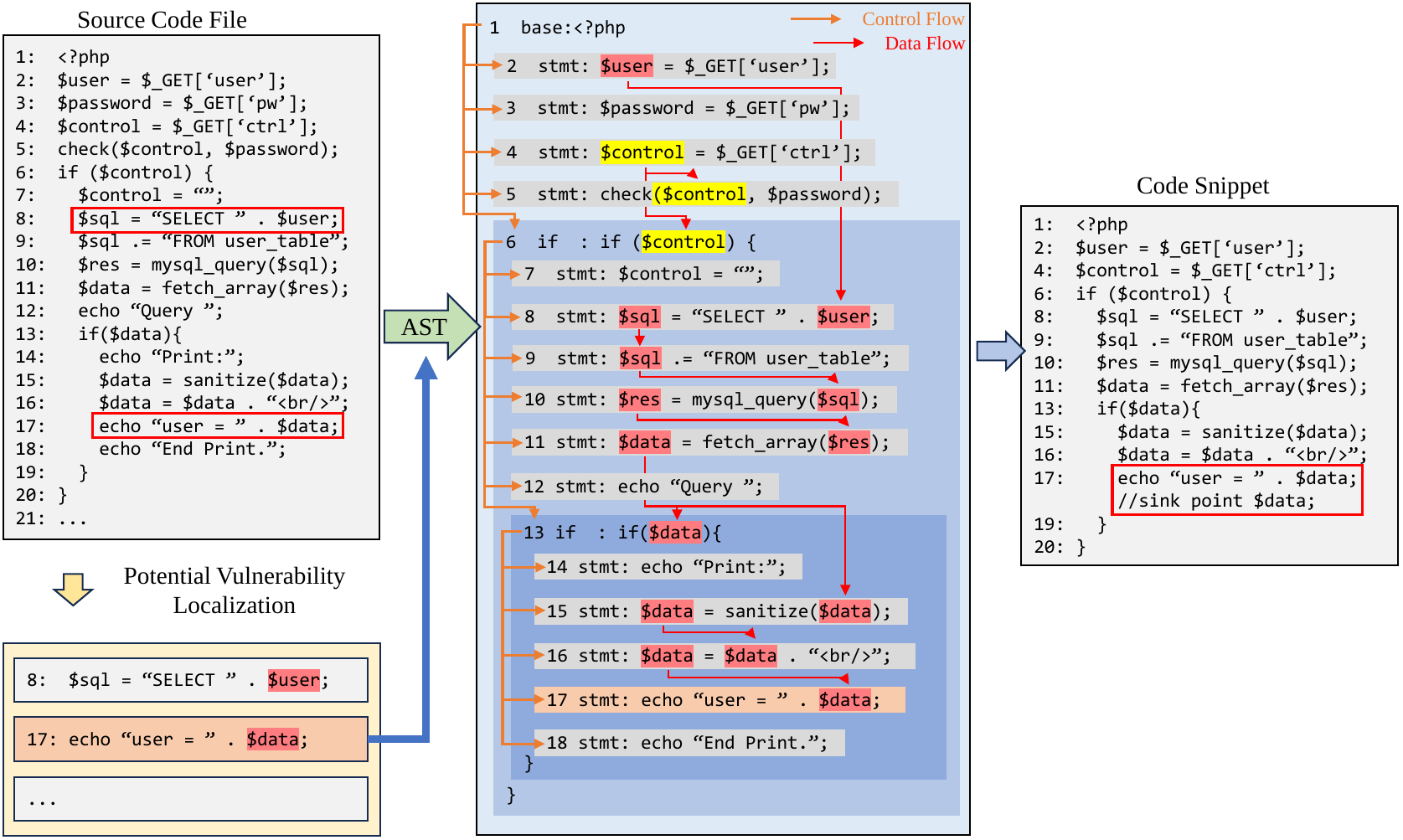}
  }
  \caption{The process of vulnerability candidate detection from a real-world PHP project. We identify potential vulnerability triggers and analyze the data flow and control flow through the source file's AST. The obtained code snippets are our samples.}
  \label{fig:vulnerability_candidate_detection}
\end{figure*}

In this phase, our objective is to scrutinize potential vulnerability triggers in the code and slice the program based on these triggers. This generates code snippets that are syntactically correct and solely associated with one vulnerability trigger, thereby reducing noise and ensuring a robust correlation between vulnerabilities and samples. Figure \ref{fig:vulnerability_candidate_detection} depicts the process of our approach of vulnerability candidate detection.

\subsubsection{Potential Vulnerability Localization}

To extract samples from source files, we initially employ our domain expertise and heuristic rule matching to identify statements in the code that could potentially trigger vulnerabilities. We have detailed specific identification methods for two types of CWE vulnerabilities in Appendix \ref{sec:appendix_Vulnerability_Candidate_Detection}.

For the identified potential vulnerability statements, we analyze the variables used for concatenation, considering each concatenated variable as a potential vulnerability source. We then mark the current variable as a tainted variable, and this variable is deemed a potential vulnerability trigger. 
The analysis of potential vulnerability triggers here is predicated on our design assumption that we already have knowledge of which functions in the program are executed freely and which code could potentially lead to vulnerabilities. In practical applications, users can modify the rules to suit their needs for more accurate matching or detecting other types of vulnerabilities.

\subsubsection{Program Slicing}\label{sec:Program Slicing}

To trim the program code, we analyze the PHP file's code to eliminate code comments, generate an Abstract Syntax Tree (AST), and extract global code, function code, and control and data flow based on the AST. We then analyze the statement where the current potential vulnerability trigger is located. Given that multiple variables in the current statement could potentially trigger the vulnerability, we replace variables other than the currently analyzed tainted variable with constants to facilitate focused analysis of a single variable. 
We search for potential vulnerability triggers related to data flow and control flow in the current analysis. We label the variables in these statements as relevant variables and further recursively search. Ultimately, we identify all statements related to potential vulnerability triggers.
While ensuring correct syntax, we extract these code statements as samples.

If the potential vulnerability is within the function, we posit that function variables are deemed untrustworthy inside the function as they are passed in from outside the function. 
Therefore, we rewrite these variables in the form of global variables "\$\_GET", and convert the function code to global code. This enables us to standardize the different representations of function code and global code. 
Upon completing the vulnerability candidate detection, we use code comments to mark taint variables at potential vulnerability triggers to enhance the sample's vulnerability representation.

\subsection{Data Preprocessing}\label{sec:Data_Preprocessing}
In this phase, our objective is to reduce irrelevant information in the samples and ensuring suitable preprocessing of the dataset. We accomplish the preprocessing of the sample set through the following three steps.

\subsubsection{Labeling}  \label{sec:Labeling}
Each extracted sample is labeled based on its potential to lead to specific types of vulnerabilities. Samples with and without vulnerabilities are labeled as $y$= \{good, bad\}.
Given that only one tainted variable is passed into the potential vulnerability statement in the samples, the labels of the samples correspond one-to-one with the variables that could potentially trigger the vulnerability. This is what we consider as the strong correlation between the samples and vulnerability information.

\subsubsection{Normalizing} 
We note that the program code contains some information that is not essential for vulnerability analysis. This information primarily includes constant strings and variable names. Due to the nature of web applications, the code often contains many constant strings, some of which are excessively long and do not significantly impact vulnerability analysis, such as strings of HTML statements. 

To preserve semantic information as much as possible, previous research often refrained from processing this content. However, in this paper, we eliminate what we consider unnecessary semantic information through keyword detection. Our experiments (sec.\ref{sec:Ablation_Study}) demonstrate that this approach is effective.

Variable names, defined by programmers, do not have a fixed form due to varying coding habits, which could potentially affect the performance of vulnerability detection models. For vulnerability analysis, the names of variables do not indicate whether the data they contain poses a threat, rendering this information unnecessary. We standardize the code by mapping identical variables to the same values and renaming different variables (i.e., "var0", "var1"). We retain user-defined function names because they provide semantic information that reveals the function's behavior.

\subsubsection{Deduplication}  
Following the normalization process, there may be similar samples in the sample set due to the removal of some irrelevant information and the presence of code reuse. Code duplication has been shown to negatively impact trained models\cite{allamanis2019codeduplication}. Ensuring the uniqueness of samples in the dataset aids the model in generalizing to the true data distribution. 
Based on sequence alignment analysis, we remove all space characters from the vulerability samples to assess their similarity and set a threshold based on experience to remove highly repetitive code.

\subsection{Model Training}
Upon the completion of sample collection and processing, we fine-tune the pre-trained code LLMs to classify the previously collected and processed samples. 
These preprocessed code samples are inputted into code LLMs, with the model's task being to ascertain the presence of vulnerabilities in the samples through sequence classification.
We employ the Low Rank Adaptive (LoRa) technique \cite{hu2021lora} to fine-tune the Query and Key in the self-attention layers of the LLM. Given the varying performance of different types of CWEs, we train models for each CWE type separately to achieve specific vulnerability detection for a particular type.

\subsection{Data Synthesis}\label{sec:data_Synthesis}
While our aim is to extract vulnerability samples through code statements that may trigger vulnerabilities, there is no existing dataset that fulfills our requirements. This is due to the fact that the vulnerability dataset of real projects originates from vulnerability repair, while synthetic datasets such as SARD do not align with the code in real-world projects. Given that existing datasets cannot meet our research needs, we designed vulnerability candidate detection and preprocessing methods to obtain datasets suitable for model training. However, manual labeling method (\ref{sec:Labeling}) limits the amount of vulnerability samples, and we need to build a large-scale dataset for model fine-tuning. Therefore, the data synthesis method introduced in this section is crucial.

Referring to data synthesis methods from existing research\cite{mirsky2023vulchecker}, which procures new vulnerability samples by inserting pure vulnerability samples into projects devoid of vulnerabilities, we extend this method to generate PHP source code samples that meets our requirements. We select samples with shorter data stream lengths and less complex conditional branches, and insert them into functions of real projects. These functions are then sliced and preprocessed to finally get our synthetic dataset. Appendix \ref{sec:appendix_Synthesis} describes the detailed process of our data synthesis.

\section{Evaluation}\label{sec:Evaluation}

This section outlines the experimental details, encompassing the experiment setup and datasets. Then we fine-tune Code LLM through data obtained from synthesis and conduct a comprehensive comparison of our method with existing methodologies, thereby validating the enhancement of our method's effectiveness and generalization capability.

\begin{table*}[t]
    \centering
    \scalebox{0.8}{
    \begin{tabular}{cccccccccccc}
    \toprule
    \multicolumn{2}{c}{\multirow{2}[2]{*}{\textbf{Methods}}} & \multicolumn{5}{c}{\textbf{CWE-79}}   & \multicolumn{5}{c}{\textbf{CWE-89}} \\
    \cmidrule(r){3-7}   \cmidrule(r){8-12} 
    & & Acc   & Rec   & Pre   & F1    &  $\Delta$F1     & Acc   & Rec   & Pre   & F1    & $\Delta$F1 \\
    \midrule
    \multirow{5}[1]{*}{\textbf{RealVul}} & CodeLlama-7b & \textbf{91.47} & \textbf{87.96} & \textbf{79.80} & \textbf{83.68} & \textcolor[rgb]{ 1,  0,  0}{+51.3} & 92.35 & 78.13 & 79.37 & 78.74 & \textcolor[rgb]{ 1,  0,  0}{+73.6} \\
          & StarCoder2-7b & 89.74 & 86.50 & 75.71 & 80.75 & \textcolor[rgb]{ 1,  0,  0}{+50.5} & 90.08 & \textbf{84.38} & 68.35 & 75.52 & \textcolor[rgb]{ 1,  0,  0}{+57.3} \\
          & StarCoder2-3b & 88.48 & 88.69 & 71.68 & 79.28 & \textcolor[rgb]{ 1,  0,  0}{+29.3} & 92.63 & 78.13 & \textbf{80.65} & \textbf{79.37} & \textcolor[rgb]{ 1,  0,  0}{+73.5} \\
          & CodeT5p-770m & 89.02 & 83.58 & 75.08 & 79.10 & \textcolor[rgb]{ 1,  0,  0}{+53.7} & 88.39 & 76.56 & 65.33 & 70.50 & \textcolor[rgb]{ 1,  0,  0}{+37.2} \\
          & CodeT5-base & 89.02 & 84.67 & 74.60 & 79.32 & \textcolor[rgb]{ 1,  0,  0}{+46.0} & 79.89 & 82.81 & 46.90 & 59.89 & \textcolor[rgb]{ 1,  0,  0}{+31.3} \\
    \midrule
    \multirow{5}[1]{*}{\textbf{Baseline}} & CodeLlama-7b & 86.51 & 26.83 & 40.74 & 32.35 &   -   & 89.52 & 3.85  & 7.69  & 5.13  & - \\
          & StarCoder2-7b & 86.51 & 24.39 & 40.00 & 30.30 &    -   & 89.80 & 15.38 & 22.22 & 18.18 & - \\
          & StarCoder2-3b & 88.27 & 48.78 & 51.28 & 50.00 &    -   & 90.93 & 3.85  & 12.50 & 5.88  & - \\
          & CodeT5p-770m & 86.22 & 19.51 & 36.36 & 25.40 &    -   & \textbf{93.20} & 23.08 & 60.00 & 33.33 &  -\\
          & CodeT5-base & 87.10 & 26.83 & 44.00 & 33.33 &   -    & 92.91 & 19.23 & 55.56 & 28.57 & - \\
    \bottomrule
    \end{tabular}%
     }
    \caption{\label{tab:on_same_sources}Evaluation results on Random Samples. $\Delta$F1 is the difference between the F1 scores of RealVul and Baseline methods.}
\end{table*}

\begin{table*}[t]
    \centering
    \scalebox{0.8}{
    \begin{tabular}{cccccccccccc}
    \toprule
    \multicolumn{2}{c}{\multirow{2}[2]{*}{\textbf{Methods}}} & \multicolumn{5}{c}{\textbf{CWE-79}}  & \multicolumn{5}{c}{\textbf{CWE-89}}  \\
    \cmidrule(r){3-7}   \cmidrule(r){8-12} 
    & & Acc   & Rec   & Pre   & F1    &   $\Delta$F1    & Acc   & Rec   & Pre   & F1    &  $\Delta$F1\\
    \midrule
    \multirow{5}[1]{*}{\textbf{RealVul}} & CodeLlama-7b & 81.71 & 75.36 & 63.41 & 68.87 & \textcolor[rgb]{ 1,  0,  0}{+43.2 } & 76.19 & \textbf{100.00} & 45.95 & 62.96 & \textcolor[rgb]{ 1,  0,  0}{+23.0 } \\
          & StarCoder2-7b & 77.43 & 63.77 & 57.14 & 60.27 & \textcolor[rgb]{ 1,  0,  0}{+56.9 } & 86.90 & 97.06 & \textbf{61.11} & \textbf{75.00} & \textcolor[rgb]{ 1,  0,  0}{+46.4 } \\
          & StarCoder2-3b & 79.76 & 75.36 & 59.77 & 66.67 & \textcolor[rgb]{ 1,  0,  0}{+41.4 } & 66.67 & 97.06 & 37.50 & 54.10 & \textcolor[rgb]{ 1,  0,  0}{+28.7 } \\
          & CodeT5p-770m & 82.88 & \textbf{73.91} & \textbf{66.23} & \textbf{69.86} & \textcolor[rgb]{ 1,  0,  0}{+19.9 } & 86.31 & 97.06 & 60.00 & 74.16 & \textcolor[rgb]{ 1,  0,  0}{+31.5 } \\
          & CodeT5-base & 74.32 & 71.01 & 51.58 & 59.76 & \textcolor[rgb]{ 1,  0,  0}{+1.1 } & 70.24 & 97.06 & 40.24 & 56.89 & \textcolor[rgb]{ 1,  0,  0}{+18.4 } \\
    \midrule
    \multirow{5}[1]{*}{\textbf{Baseline}} & CodeLlama-7b & 89.38 & 17.86 & 45.45 & 25.64 & -     & \textbf{87.23} & 37.84 & 42.42 & 40.00 & - \\
          & StarCoder2-7b & 89.74 & 1.79  & 33.33 & 3.40  & -     & 86.32 & 24.32 & 34.62 & 28.57 & - \\
          & StarCoder2-3b & 89.19 & 17.86 & 43.48 & 25.32 & -     & 85.71 & 21.62 & 30.77 & 25.40 & - \\
          & CodeT5p-770m & 89.01 & 53.57 & 46.88 & 50.00 & -     & 84.50 & 51.35 & 36.53 & 42.70 & - \\
          & CodeT5-base & \textbf{89.93} & 69.64 & 50.65 & 58.65 & -     & 85.41 & 40.54 & 36.58 & 38.46 & - \\
    \bottomrule
    \end{tabular}
    }
    \caption{\label{tab:on_different_sources}Evaluation results on Unseen Projects. $\Delta$F1 is the difference between the F1 scores of RealVul and Baseline methods.}
\end{table*}

\subsection{Experiments and Datasets}\label{sec:Experiments_and_Datasets}

\subsubsection{Experiments}\label{sec:Experiments}

We execute extensive experiments to validate the performance of RealVul, considering two scenarios: model prediction code and model training code derive from identical and different PHP projects. We design three related experiments accordingly.

\textbf{EXP1: Effectiveness}. For the first scenario, we do not distinguish the projects where the samples in the training, validation, and test sets come from. As a baseline, We compare the performance of RealVul and LLMs fine-tuned by vulnerability-repair-based datasets. These dataset will be randomly divided into training, validation and test sets.


\textbf{EXP2: Generalization}. For the second scenario, We require that the data used for testing is unknown to the fine-tuned LLMs. This means that the training set, validation set and test set are as unrelated as possible in terms of data sources. For the dataset of baseline method, we impose the same requirements, ensuring the these three sets are sourced from different projects.


To further demonstrate the real-world application capabilities of RealVul, we conduct a comparison of RealVul with two common PHP SAST tools, RIPS\cite{RIPS} and Fortify SCA\cite{fortify}, in terms of their function-level analysis capabilities.

\textbf{EXP3: Ablation Study}. Normalization and model training are two crucial component of our sample processing. To demonstrate that normalization generally enhances model performance, we conduct corresponding experiments on CWE-79 vulnerabilities. We use non-normalized datasets for training and testing, and compare the results with the first two experiments. We also compare RealVul with in-context learning approaches to demonstrate the necessity of fine-tuning.  We use both Zero-shot and Few-shot prompts. 


In line with the evaluation metrics of existing researches\cite{chakraborty2020Reveal, fu2022linevul, chen2023diversevul}, we employ four metrics: accuracy, recall, precision, and F1 score to thoroughly evaluate our experimental results. During the sampling phase, we implement AST generation of PHP code using the PHPLy \cite{phply}. We primarily use CodeT5\cite{wang2021codet5}, CodeT5p\cite{wang2023codet5+}, CodeLlama\cite{roziere2023codellama}, and the latest StarCoder2\cite{lozhkov2024starcoder} to demonstrate the effectiveness of RealVul, including five models: CodeT5-base, CodeT5p-770m, CodeLlama-7b, StarCoder2-3b, and StarCoder2-7b. 
In the Ablation Study on model fine-tuning, we utilized the in-context learning approaches to directly evaluate GPT-4, as well as other SOTA open-source LLMs such as Mistral-7B, Llama3-8B, and CodeLlama-7B. We used both zero-shot and few-shot prompts. In the few-shot prompts, we provided two demonstration examples to help the LLMs understand the context of the task. The Appendix \ref{sec:appendix_Experiments_Setup} provides detailed information about models and evaluation metrics.

\subsubsection{Datasets}\label{sec:Datasets}
We collect samples from the CrossVul dataset\cite{nikitopoulos2021crossvul} for training and testing models of RealVul and baseline method. The CrossVul dataset provides PHP files pre and post vulnerability repair, which allows us to collect the datasets from the same data source in different ways. We use the RealVul framework to obtain real-world vulnerability datasets, denoted as $D_{real}$. Leveraging our data synthesis algorithm, we synthesize a large-scale dataset $D_{syn}$ from $D_{real}$ and the SARD dataset\cite{sard}. For the baseline method, we obtaine another dataset by comparing the code differences pre and post vulnerability repair, denoted as $D_{rep}$.

To evaluate RealVul, we split the $D_{syn}$ as the training and validation sets, and we use $D_{real}$ as the test set. For the baseline method, we split the $D_{syn}$ as the training, validation and test sets. In \textbf{EXP1}, we split the dataset through random sampling. But in \textbf{EXP2}, we strictly split the datasets based on the source projects. We stipulate that the code samples in the test set can't be used for synthesizing training and validation sets, and the samples used for synthesizing validation sets could not be used for synthesizing training sets. In \textbf{EXP3}, we use non-normalized datasets $D_{real}$* and $D_{syn}$* for training and testing, and we use $D_{real}$ to evaluate the in-context learning approaches. We provide more information in Appendix \ref{sec:appendix_Experiments_Setup}, including statistics and properties of our datasets.

\subsection{Effectiveness}

Table \ref{tab:on_same_sources} presents the evaluation outcomes of two vulnerabilities, CWE-79 (XSS) and CWE-89 (SQLI), utilizing the same data source on our test set $D_{real}$. Based on these results, we can infer the following:

Despite our training data being algorithmically synthesized, the evaluation outcomes of RealVul exhibit commendable performance across four metrics. Even for the CWE-89 type, which has a smaller real data sample size, the evaluation outcomes remain relatively stable. This suggests that our synthesized dataset aligns well with the real dataset, and training with synthesized data can effectively evaluate code samples in real environments. Due to the limited data volume in CWE-89, its overall F1 score performance is not as stable as that of CWE-79, indicating potential for further enhancement of our method's analytical capability by increasing the number of real sample data.

Comparing with the baseline method further reveals that our RealVul method generally outperforms the baseline method, with RealVul combined with CodeLlama-7b and StarCoder2-3b delivering the best performance in the CWE-79 and CWE-89 tasks, particularly in terms of F1 score. In contrast, the F1 score of the vulnerability repair-based sampling method does not exceed 50\%. This suggests that our sampling and processing techniques enable our code to better represent vulnerability feature information, thereby enhancing the LLM code's performance in vulnerability detection.

\subsection{Generalization}

Table \ref{tab:on_different_sources} displays the evaluation outcomes on test sets from different data sources. These results substantiate that our RealVul method more effectively encapsulates vulnerability-related information in the code and achieves superior generalization performance.

In our training data, the proportion of vulnerability samples is relatively small, reflecting the uneven distribution of vulnerability samples in real environments and imposing higher demands on the analytical method's capability. Therefore, although our method may slightly lack accuracy compared to the baseline, our F1 score significantly surpasses the baseline. The baseline's input is complete function code or top-level code, which increases the model's analytical difficulty compared to samples obtained by our method. 

Additionally, we observe that the model's parameter count has minimal impact on its vulnerability detection capability. Even smaller models like Codet5 and Codet5p possess sufficient analytical capabilities to accomplish the analysis task. This suggests that our method, by vulnerability candidate detection and preprocessing, reduces individual sample code length and emphasizes potential vulnerability trigger-related information, thereby reducing the model's analytical capability requirements.

\begin{table}[t]
    \centering
    \setlength{\tabcolsep}{0.78mm}
    \scalebox{0.81}{
    \begin{tabular}{cccccccc}
    \toprule
    \multicolumn{2}{c}{\multirow{2}[2]{*}{\textbf{Methods}}} & \multicolumn{3}{c}{\textbf{CWE-79}} & \multicolumn{3}{c}{\textbf{CWE-89}} \\
    \cmidrule(r){3-5}  \cmidrule(r){6-8} 
    &   & TP    & FP    & Times (s) & TP    & FP    & Times (s) \\
    \midrule
    \multirow{5}{*}{\textbf{RealVul}} & CodeLlama-7b & 40 & 14    & 152   & \textbf{30} & 17    & 147 \\
    & StarCoder2-7b & 34    & 24    & 130   & 29 & 22    & 151 \\
    & StarCoder2-3b & 38    & 11 & 56    & 29 & 9 & 64 \\
    & CodeT5p-770m & 39    & 15    & 23    & 29 & 20    & 32 \\
    & CodeT5-base & 32    & 16    & 7     & 29 & 12     & 10 \\
    \midrule
    \multicolumn{2}{c}{\textbf{RIPS}} & \textbf{43}    & 30    & $<$1   & 3     & 3     & $<$1 \\
    \multicolumn{2}{c}{\textbf{Fortify SCA}} & 40    & \textbf{8}    & 30    & 1     & \textbf{0}     & 28 \\
    \bottomrule
    \end{tabular}
    }
    \caption{\label{tab:on_SAST} Comparison of RealVul and two SAST tools. We also provide the time required for the evaluation.}
\end{table}

We also present the comparative results of RealVul and the two traditional SAST tools in Table \ref{tab:on_SAST}. From the results, our method performs slightly worse than static tools on CWE-79. However, our method obviously outperforms static tools on CWE-89. This is because we match the behavior of concatenating SQL statements rather than the functions executing SQL statements when selecting potential vulnerability points for SQL injection vulnerabilities. This allows us to more comprehensively identify SQL vulnerabilities.

In traditional methods, the improvement of vulnerability detection requires the continuous accumulation of rules, which increases the time necessary for analysis. It is worth noting that while SAST tools primarily rely on the CPU for computation during runtime, the part of our method that applies the CodeT5 series models takes less evaluation time than SAST tools, and the accuracy is fairly close to that of traditional SAST tools. 
In Appendix \ref{sec:appendix_Results}, we provide further explanation through case study.

\begin{figure}[t]
    \centering
    \subfigure[Test on Random Samples] {\includegraphics[width=.45\textwidth]{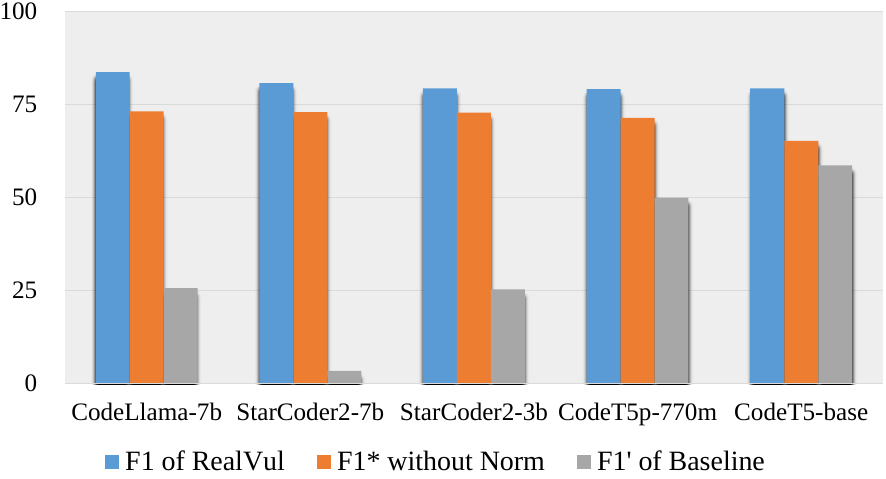}}
    \subfigure[Test on Unseen Projects] {\includegraphics[width=.45\textwidth]{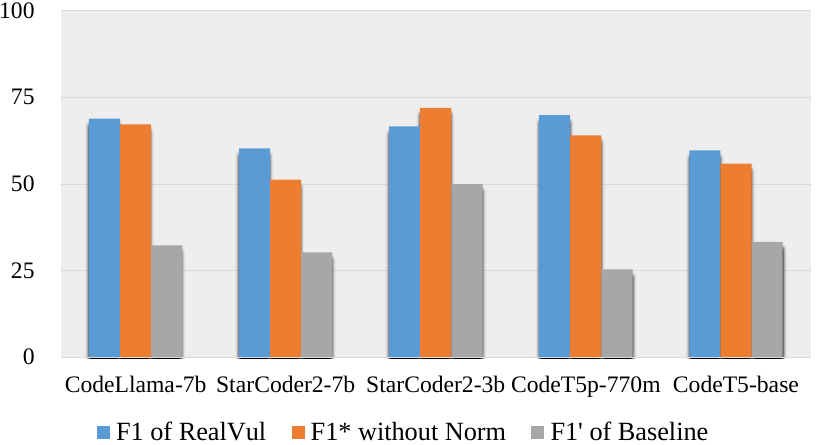}}

    \caption{Comparison of ablation study results with the visualization of results from the first two experiments.}
    \label{fig:ablation_study_detail}
\end{figure}

\subsection{Ablation Study}\label{sec:Ablation_Study}

The ablation study results of normalization, presented in Figure \ref{fig:ablation_study_detail}, clearly show that although StarCoder2-3b experiences a certain decrease in F1 scores when tested on unseen projects, normalization processing is necessary in most cases, with the maximum F1 score difference reaching 14.06\%. This suggests that appropriately reducing irrelevant semantic information may lead to better results in PHP vulnerability detection tasks.
What's more, normalizing the function name might yield better results. However, to retain the function's necessary information, it is essential to construct a function call graph and analyze its functionality, which necessitates further development of an analysis system.

Although the absence of normalization processing can lead to a decrease in evaluation outcomes, our sampling method still outperforms Baseline in previous experiments, indicating its superiority over vulnerability repair-based methods. We present all the evaluation outcomes of this ablation study in Appendix \ref{sec:appendix_Results}.

The evaluation results of ablation study on model fine-tuning are shown in Table \ref{tab:ablation_study_on_fine-tuning} in Appendix \ref{sec:appendix_Results}. It is obvious that there is a significant performance gap between the in-context learning approaches and RealVul, which shows the contribution of our RealVul. This also confirms the conclusion of existing research\cite{steenhoek2024comprehensive} that the models lack the ability to directly use in-context learning to understand software vulnerabilities, thus fine-tuning is necessary.

\section{Conclusion}

In this paper, we concentrate on PHP web vulnerability detection utilizing code LLMs. 
To enhance detection proficiency, we identifies potential vulnerability triggers, analyzes control flow and data flow, and eliminates unnecessary semantic information to obtain samples robustly correlated with vulnerability information. Based on the improved data synthesis method, we extensively synthesize new vulnerability samples, thereby alleviating the challenge of insufficient vulnerability dataset of PHP.
We carry out extensive experiments, comparing our method with existing techniques using samples derived from real-world projects. The experimental outcomes indicate that our method exhibits significantly superior capabilities in comparison to existing techniques.

\section*{Limitations}

We acknowledge three potential limitations in our study that warrant further exploration in future research: (i) During the normalization process, we preserve user-defined function names. We posit that integrating function call analysis, standardizing function name representation, and supplementing code comments with functional information could potentially enhance the effectiveness of our approach. (ii) We fine-tune our model specifically for each type of CWE vulnerability to augment detection capabilities. This approach incurs substantial overhead and the performance of a unified multi-classification model merits investigation. (iii) At present, there is a dearth of effective vulnerability sample labeling methods, leading us to resort to manual labeling of samples and data synthesis to compensate for the lack of sufficient data volume. Consequently, the efficacy of our dataset could be further improved. (iiii) We adopted a heuristic rule-based approach to identify potential vulnerability triggering statements, and these rule bases still need to be extended or modified in our framework to adapt to new vulnerability patterns or updated programming practices.

\section*{Acknowledgments}
This work is supported by the National Key Research and Development Program of China (2022YFB3105405, 2021YFC3300502) and the Provincial Key Research and Development Program of Anhui(202423l10050033).

\bibliography{custom}

\appendix

\section{Related Works}\label{sec:appendix_Related}
The statistical data of existing vulnerability detection datasets and research methods are shown in Tables \ref{tab:vulnerability_datasets} and \ref{tab:exsiting_approches}.

\begin{algorithm}[t]
\caption{Data Synthesis}
\fontsize{9.5pt}{11.5pt}\selectfont 
\label{alg:Data Synthesis}
\begin{algorithmic}
\STATE \textbf{Input:} Existing pure sample set $S_{raw}$, projects' global code and function code set $C_{proj}$ and synthesis times $T$ for each sample,
\STATE \textbf{Output:} Synthesis samples set $S_{syn}$ obtained through synthesis.
\STATE $S_{syn} \gets \emptyset$
\FOR{ each $s_{raw} \in S_{raw}$ }
    \STATE $C_{raw} \gets$ code list of Top-level AST nodes in $s_{raw}$
    \FOR{ each $c_{proj} \in C_{proj}$ }
        \FOR{ $i$ in range $T$ }
            \STATE $c_p \gets$ code of random path in the control flow of $c_{proj}$ 
            \STATE $c_p \gets$ \textbf{remove\_vuln\_triggers}($c_p$)
            \STATE $ c_{syn} \gets$ Randomly insert $C_{raw}$ into $c_p$
            \IF{\textbf{syntax\_check}( $ c_{syn} $ )}
                \STATE $s_{syn} \gets $\textbf{slicing\_and\_preprocessing}( $c_{syn}$)
                \STATE $S_{syn} \gets S_{syn} \cup \{s_{syn}\}$
            \ENDIF
        \ENDFOR
    \ENDFOR
\ENDFOR
\end{algorithmic}
\end{algorithm}

\begin{table*}[t]
    \centering
    \setlength{\tabcolsep}{2mm}
    \scalebox{0.9}{
    \begin{tabular}{llll}
    \toprule
         \textbf{Datasets} &  \textbf{Size} &  \textbf{Type} &  \textbf{Language} \\ 
    \midrule
        SARD\cite{sard} & 450K & Synthetic & C/C++/Java/PHP/C\# \\ 
        SATE IV Juliet\cite{juliet2013} & 253K & Synthetic & C/C++/Java \\ 
        Devign\cite{zhou2019devign} & 23K & Real & C \\ 
        Reveal\cite{chakraborty2020Reveal} & 18K & Real & C \\ 
        Big-Vul\cite{fan2020Bigvul} & 188K & Real & C/C++ \\ 
        D2A\cite{D2A} & 1.3M & Real & C/C++ \\ 
        CrossVul\cite{nikitopoulos2021crossvul} & 27K & Real & C/C++/Java/PHP/... \\ 
        DiverseVul\cite{chen2023diversevul} & 349K & Real & C/C++ \\ 
    \bottomrule
    \end{tabular}
    }
    \caption{\label{tab:vulnerability_datasets}Overview of Existing Vulnerability Datasets}
\end{table*}

\begin{table*}[t]
    \centering
    \setlength{\tabcolsep}{2mm}
    \scalebox{0.9}{
    \begin{tabular}{llll}
    \toprule
         \textbf{Approaches} &  \textbf{Input} &  \textbf{Model} & \textbf{Language} \\
    \midrule
        VulDeePecker\cite{li2018vuldeepecker} & Token (Code Gadget) & Bi-LSTM & C/C++ \\ 
        $\mu$VulDeePecker\cite{zou2019muVulDeePecker} & Token (Code Gadget) & B-BLSTM & C/C++ \\ 
        Devign\cite{zhou2019devign} & Graph (CPG) & GCN & C \\ 
        Reveal\cite{chakraborty2020Reveal} & Graph (CPG) & GGNN & C \\ 
        SySeVR\cite{li2021sysevr} & Token (SeVC) & BRNN, BGRU & C/C++ \\ 
        \makecell[l]{LineVul\\\cite{fu2022linevul}} & Token (Code) & CodeBert & C/C++ \\ 
        LineVD\cite{hin2022linevd} & Graph (PDG) & GCN, GAT & C/C++ \\
        VulChecker\cite{mirsky2023vulchecker} & Graph (ePDG) & S2V, DNN & C/C++ \\ 
        DiverseVul\cite{chen2023diversevul} & Token (Code) & \makecell[l]{RoBERTa, GPT-2,\\ CodeT5} & C/C++ \\ 
        RealVul(Ours)& Token (Processed Code) & \makecell[l]{CodeT5, CodeT5+, \\CodeLlama, StarCoder2} & PHP \\
    \bottomrule
    \end{tabular}
    }
    \caption{\label{tab:exsiting_approches}Overview of Existing Vulnerability Detection Approches}
\end{table*}

\begin{table*}[t]
    \centering
    \setlength{\tabcolsep}{2.2mm}
    \scalebox{0.9}{
    \begin{tabular}{cccccc}
    \toprule
    \textbf{Code LLM} & \textbf{Size} & \textbf{Release Time} & \textbf{Base Model} & \textbf{Publisher} & \textbf{License}\\
    \midrule
    CodeT5-base & 220M  &  Sep-2021 & T5    & Salesforce &  Open-source
\\
    CodeT5p-770m & 770M  &  May-2023 & T5    & Salesforce &  Open-source\\
    CodeLlama-7b & 7B    &  Jun-2023 & Llama2-7b & Meta AI &  Open-source\\
    StarCoder2-3b & 3B    &  Feb-2024 & -     & BigCode &  Open-source\\
    StarCoder2-7b & 7B    &  Feb-2024 & -     & BigCode &  Open-source\\
    Llama3-8b-instruct & 8B    &  apr-2024 & - & Meta AI &  Open-source\\
    Mistral-7b-instruct-v0.3 & 7B    &  Dec-2023 & Mistral-7b     & Mistral AI &  Open-source\\
    GPT-4 & -    &  Mar-2023 & -     & OpenAI &  Closed-source\\
    \bottomrule
    \end{tabular}
    }
    \caption{\label{tab:model_detail} Detail information of Models we apply in this paper.}
\end{table*}

\section{Vulnerability Candidate Detection}\label{sec:appendix_Vulnerability_Candidate_Detection}
Different methods are needed to identify potential vulnerability triggers for different types of vulnerabilities. For XSS vulnerabilities (CWE-79), the potential statement is \emph{echo}, \emph{print} and other output statements.

However, searching for functions that execute SQL statements through matching as potential vulnerability points is highly inaccurate. On the one hand, many PHP projects do not use PHP's native SQL execution functions, and they often use self written SQL execution functions or some PHP framework functions, which makes it difficult to match all potential vulnerability points. On the other hand, the matched SQL execution function cannot determine whether a filter for SQL injection vulnerabilities is built-in. In practice, SQL statements often need to be concatenated with variables before execution, so we match statements that directly concatenate SQL statements with variables as potential vulnerability trigger statements.

\section{Data Synthesis}\label{sec:appendix_Synthesis}

Algorithm \ref{alg:Data Synthesis} demonstrates our data synthesis method. By analyzing the AST of the samples, we use the code corresponding to the top-level AST node as the basic unit for inserting into the clean project code. Subsequently, we analyze the control flow of the project code, randomly select a control flow path, and remove potential code trigger statements to obtain clean project code. After T rounds of synthesis, syntax checking of the code, program slicing, and preprocessing as described in sections \ref{sec:Vulnerability_Candidate_Detection} and \ref{sec:Data_Preprocessing}, we ultimately obtain our synthesized dataset.

\begin{table}[t]
    \centering
    \begin{tabular}{c}
    \multicolumn{1}{c}{(a): Test on Random Samples} \\
        \scalebox{0.85}{
        \begin{tabular}{ccccc}
        \toprule
            \multirow{2}{*}{\textbf{Code LLM}} & \multicolumn{4}{c}{\textbf{Metrics}}\\
            \cmidrule(r){2-5} 
            & \textbf{Acc} & \textbf{Rec} & \textbf{Pre} & \textbf{F1}\\
            \midrule
            CodeLlama-7b & 83.75 & 89.05 & 62.09 & \textbf{73.16} \\
            StarCoder2-7b & 83.30 & \textbf{90.51} & 61.08 & 72.94 \\
            StarCoder2-3b & 83.48 & 89.05 & 61.62 & 72.83 \\
            CodeT5p-770m & \textbf{84.66} & 77.01 & \textbf{66.56} & 71.40 \\
            CodeT5-base & 82.03 & 67.88 & 62.83 & 65.26 \\
            \bottomrule
        \end{tabular} } \\ 
        \\
    \multicolumn{1}{c}{(b): Test on Unseen Projects} \\
        \scalebox{0.85}{
        \begin{tabular}{ccccc}
        \toprule
             \multirow{2}{*}{\textbf{Code LLM}} & \multicolumn{4}{c}{\textbf{Metrics}}\\
            \cmidrule(r){2-5} 
            & \textbf{Acc} & \textbf{Rec} & \textbf{Pre} & \textbf{F1}\\
            \midrule
            CodeLlama-7b & \textbf{84.04} & 60.87 & \textbf{75.00} & 67.20 \\
            StarCoder2-7b & 77.04 & 44.93 & 59.62 & 51.24 \\
            StarCoder2-3b & 80.93 & \textbf{91.30} & 59.43 & \textbf{72.00} \\
            CodeT5p-770m & 74.71 & 84.05 & 51.78 & 64.08 \\
            CodeT5-base & 75.48 & 57.97 & 54.05 & 55.94 \\
            \bottomrule
        \end{tabular}  }%
      
    \end{tabular}
    \caption{\label{tab:ablation_study_detail} Evaluation results of Ablation Study on normalization.}
\end{table}

\section{Experiments Setup}\label{sec:appendix_Experiments_Setup}

\begin{table*}[t]
    \centering
    \setlength{\tabcolsep}{2.2mm}
    \scalebox{0.85}{
    \begin{tabular}{cccccccccc}
    \toprule
    \multirow{2}[2]{*}{\textbf{CWE}} & \multirow{2}[2]{*}{\textbf{\# Projects}} & \multicolumn{2}{c}{\textbf{\# by Fix ($D_{rep}$)}} & \multicolumn{2}{c}{\textbf{\# RealVul ($D_{real}$)}} & \multicolumn{2}{c}{\textbf{\# for Synthesis}} & \multicolumn{2}{c}{\textbf{ \# Synthesis ($D_{syn}$)}} \\
    \cmidrule(r){3-4}  \cmidrule(r){5-6} \cmidrule(r){7-8} \cmidrule(r){9-10}    
    & & Total & Vuln  & Total & Vuln  & Total & SARD  & Total & Vuln \\
    \midrule
    \textbf{CWE-79} & 154   & 6818  & 815   & 1102  & 274   & 1417  & 315   & 33255 & 12040 \\
    \textbf{CWE-89} & 50    & 3525  & 303   & 353   & 64    & 543   & 190   & 14116 & 4237 \\
    \textbf{Total} & 180   & 10343 & 1118  & 1455  & 338   & 1960  & 505   & 47371 & 16277 \\
    \bottomrule
    \end{tabular}
    }
    \caption{\label{tab:dataset_detail}Statistics of the dataset we used. We list the number of samples obtained through vulnerability repair, samples obtained through RealVul, samples used for data synthesis, and samples obtained through synthesis.}
\end{table*}

\begin{table*}[ht]
    \centering
    \scalebox{0.9}{
    \begin{tabular}{cccccccccc}
    \toprule
    \multicolumn{2}{c}{\multirow{2}[2]{*}{\textbf{Methods}}} & \multicolumn{4}{c}{\textbf{CWE-79}}  & \multicolumn{4}{c}{\textbf{CWE-89}}  RVul\\
    \cmidrule(r){3-6}   \cmidrule(r){7-10} 
    & & Acc   & Rec   & Pre   & F1        & Acc   & Rec   & Pre   & F1    \\
    \midrule
    \multirow{1}[1]{*}{\textbf{RealVul}} & CodeLlama-7b & 91.47 & 87.96 & 79.80 & \textbf{83.68}  & 92.35 & 78.13 & 79.37 & \textbf{78.74}\\
    \midrule
    \multirow{4}[1]{*}{\textbf{zero-shot}} & CodeLlama-7b & 24.86 & 100.00 & 24.86 & 39.82    & 18.18 & 	100.00 & 18.18 & 30.77  \\
          & Llama3-8b & 24.86 & 100.00  & 24.86 & 39.82      & 18.13 & 100.00 & 18.13 & 30.69  \\
          & Mistral-7b & 24.01 & 100.00 & 24.01	 & 38.72     & 17.69 & 100.00 & 17.09 & 29.19  \\
          & GPT-4 & 27.58 & 100.00 & 25.56 & 40.71   & 18.69 & 100.00 & 18.23 & 30.84  \\
          \midrule
    \multirow{4}[1]{*}{\textbf{few-shot}} & CodeLlama-7b & 24.58 & 97.36 & 21.63 & 	35.40    & 10.81 & 100.00 & 10.20 & 18.52  \\
          & Llama3-8b & 25.17 & 100.00  & 25.17 & 40.22      & 15.04 & 100.00 & 15.04 & 26.15  \\
          & Mistral-7b & 33.48 & 97.44 & 26.88 & 42.14     & 24.36 & 89.06 & 17.98 & 29.92  \\
          & GPT-4 & 59.50 & 96.61 & 41.91 & \textbf{58.46}   & 40.79 & 98.43 & 23.24 & \textbf{37.61}  \\
    \bottomrule
    \end{tabular}
    }
    \caption{\label{tab:ablation_study_on_fine-tuning}Evaluation results of Ablation Study on model fine-tuning.}
\end{table*}

\begin{figure*}[htb]
    \centering
    \scalebox{1}{
    \subfigure[CWE-79 Case] {
        \begin{minipage}{0.45\textwidth}
            \centering
            \includegraphics[width=\textwidth]{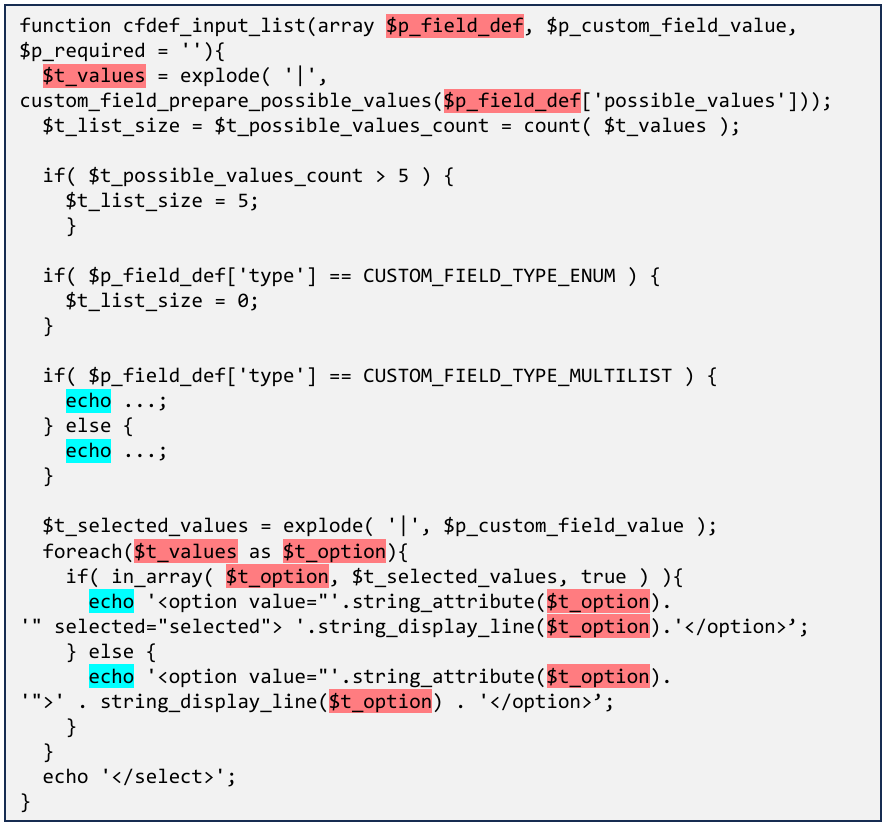} \\
            \hfill
            \hfill
            \includegraphics[width=\textwidth]{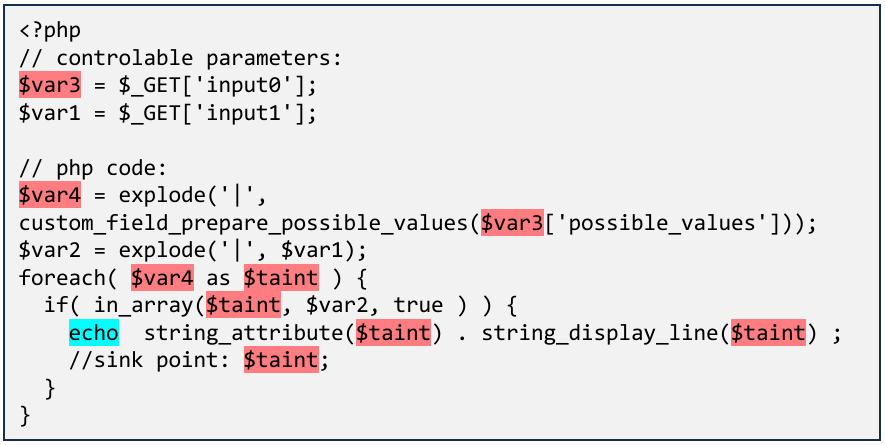}
        \end{minipage}
    }%
    \subfigure[CWE-89 Case] {\label{fig:89case}
        \begin{minipage}{0.45\textwidth}
            \centering
            \includegraphics[width=\textwidth]{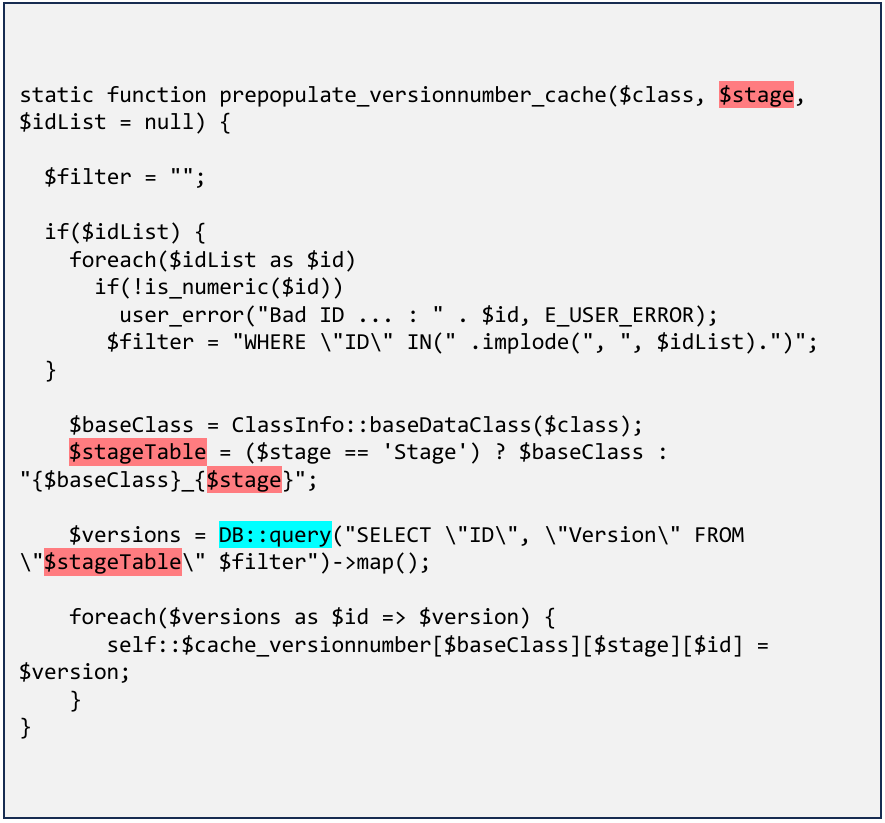} \\
            \includegraphics[width=\textwidth]{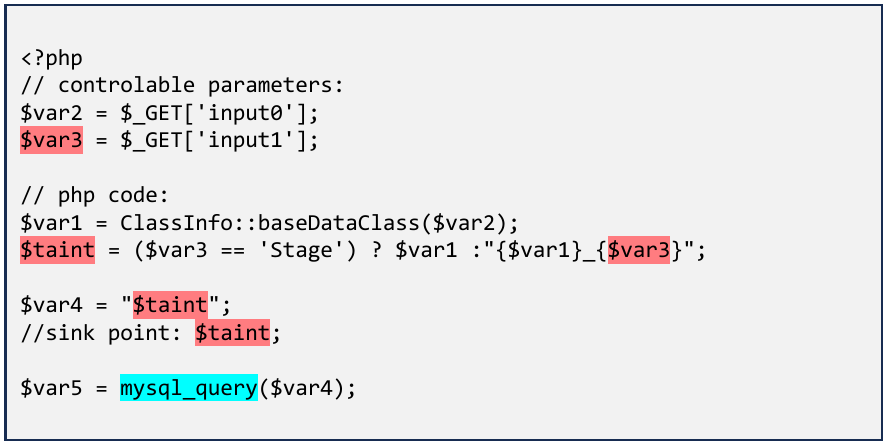}
        \end{minipage}
    }%
    }
    \caption{Two sets of sample Cases obtained through vulnerability reapir and RealVul. We mark the data flow and potential vulnerability statements.}
    \label{fig:Case_Study}
\end{figure*}

\textbf{Model Configuration}. Table \ref{tab:model_detail} presents an overview of the Code LLMs we apply in our paper. We use cross entropy as the loss function and deploy LLMs on four NVIDIA-V100 GPU with 32GB of memory for training and testing to demonstrate the effectiveness of RealVul. We adopt Adam optimizer in fp16 precision, 32 global batch size. We set the training epoch to 2 and 3 for test on random samples and test on unseen projects. We add an additional epoch for samples of CWE-89 because its data is small.

\textbf{Datasets}. Table \ref{tab:dataset_detail} presents an overview of the dataset utilized in our paper. The sample size derived from our methodology is comparatively smaller than that obtained through vulnerability repair, primarily due to the constraints of our manual labeling process. Furthermore, we pinpoint potential triggers for CWE-89 type vulnerabilities by scrutinizing instances where variables are amalgamated into SQL statements within the code. Despite drawing from the vulnerability dataset, such SQL concatenation instances are relatively infrequent in comparison to "echo" and "print" statements, leading to a reduced collection of CWE-89 vulnerability samples. To counteract this issue, we amplified the frequency of single sample synthesis ($T$) during the data synthesis process, resulting in synthesized samples constituting approximately 30\% of the total sample size. This strategy has somewhat alleviated the problem.

Our datasets are at the snippet level, and the average LoC for the XSS and SQL vulnerability datasets are 13.07 and 22.06 lines, respectively. Compared with function-level datasets, the advantages of our snippet-level dataset are fine-grained detection, simple labeling, more samples, high training efficiency, and strong sample independence. However, the contextual information provided by ours is relatively limited compared to function-level and project-level. 

\textbf{Evaluation Metrics}. We use the following evaluation metrics:
\begin{itemize}
    \item Accuracy indicates the overall correctness: $Acc = \frac{TP + TN}{TP + FP + FN + TN}$.
    \item Precision indicates the correct positive predictions part: $Pre = \frac{TP}{TP + FP}$.
    \item Recall calculates the correctly recalled positive examples part: $Rec = \frac{TP}{TP + FN}$.
    \item F1 is the harmonic mean of Precision and Recall: $F1 = 2 \cdot \frac{Pre \cdot Rec}{Pre + Rec}$. We mainly use F1 to decide the best performing model as it provides a balanced evaluation of the model's performance in terms of both Precision and Recall.
\end{itemize}

\section{Experiment Results}\label{sec:appendix_Results}

\textbf{Case Study}. We illustrate two sample cases in the Figure \ref{fig:Case_Study}. In practical projects, there may exist longer top-level or function codes. It can be observed that compared to the original samples, our samples are shorter, contain fewer irrelevant details and maintain correct syntax. This enables the Code LLMs to more easily perform vulnerability detection tasks. Additionally, as the function code shown in the Figure \ref{fig:89case}, triggering functions in CWE-89 vulnerabilities are difficult to identify through traditional rules. That's why our potential vulnerability localization method significantly outperforms SAST tools in CWE-89 vulnerability detection.

\textbf{Ablation Study}. The detailed evaluation results of ablation study on normalization and model fine-tuning are shown in the Table \ref{tab:ablation_study_detail} and Table \ref{tab:ablation_study_on_fine-tuning}.

\end{document}